\documentclass[]{spie}  

\usepackage{amsmath,amsfonts,amssymb}
\usepackage{graphicx}
\usepackage[colorlinks=true, allcolors=blue]{hyperref}

 \newcommand{\V}[1]{\boldsymbol{#1}}      
 \newcommand{\M}[1]{\mathbf{#1}}          
 \newcommand{\T}{^\mathrm{T}}             

\newcommand{\norm}[1]{\Vert #1\Vert}

\newcommand{\avg}[1]{\langle #1\rangle}

\newcommand{\Cerror}{\M{C}_{\V{e}}}      
\newcommand{\Cprior}{\M{C}_{\V{w}}}      

\title{Overview of multi-conjugate adaptive optics reconstructors}

\author[a]{Cl{\'e}mentine  B{\'e}chet}
\affil[a]{Institute of Mathematical and Computational Engineering, Pontificia Universidad Cat{\'o}lica de Chile, Santiago, Chile}

\authorinfo{Further author information: (Send correspondence to C. B{\'e}echet)\\ C. B{\'e}chet: E-mail: cbechetp@ing.puc.cl, Telephone: +56 2354 9597}

\begin{document}
\maketitle

\begin{abstract}
  Multi-Conjugate Adaptive Optics (MCAO) systems aim at correcting for
  the atmospheric turbulence uniformly over wide-field
  observations. Compared to classical AO, this requires a tomographic
  reconstruction of the turbulence perturbations, that is to say a
  reconstruction of the atmosphere in its volume, and the control of
  several deformable mirrors optically conjugated to various
  altitudes.

  During the last two decades, many tomographic reconstructors have
  been proposed in the AO literature and some have been implemented on
  sky on existing MCAO systems. This paper presents an overview of
  MCAO reconstructors. Four categories of reconstructors are
  considered: i) the reconstructors based on interaction matrix; ii)
  the minimum-variance static reconstructors; iii) the 3-step
  recontructors and iv) the Kalman-based reconstructors. We discuss
  their relative advantages and drawbacks.

\end{abstract}


\section{INTRODUCTION}
\label{sec:intro}  

Classical adaptive optics (AO) systems only require to estimate the
resulting wavefront distortions in the pupil plane. These distortions
are then compensated by controlling a single deformable mirror
optically conjugated to this pupil plane. In such configuration, the
AO correction progressively decorrelates with the resulting
atmospheric disturbance when observing at larger separation
angles. This is because the atmospheric turbulence is distributed in a
volume between the telescope and approximately 20~km above the
aperture and the two light paths through the atmosphere differ. When
the AO guide star is too far off-axis with respect to the science
observation direction, the residuals of turbulence become too
significant to provide satisfying AO correction. This effect is called
the anisoplanatism \cite{Fried1982a} of the atmospheric turbulence in
AO. This limitation is characterized by the isoplanatic patch angular
size of the atmospheric turbulence \cite{Fried1982a}, $\theta_0$. If
the angle between the guiding and the observing direction is increased
by $\theta$, the AO correction is degraded by a factor
$\exp[- (\theta/\theta_0)^{5/3}]$. Typical isoplanatic patch values
$\theta_0$ are of the order of $2-3''$ at 500~nm and $10-20''$ in
K-band (at $2.2\mu$m). This means that the classical AO correction is
limited to small fields around a star bright enough to be used as
reference.

The goal of multi-conjugate adaptive optics (MCAO)\cite{Beckers1989a}
is thus to beat the anisoplanatism effect and provide uniform AO
correction over wide-field observations. To allow AO over a wider
field-of-view (FOV), the wavefront sensing must be done in various
directions simultaneously among this FOV. From the combination of
these data, a tomographic reconstruction of atmospheric disturbances
is made. Such reconstruction aims to disentangle the heights at which
wavefront distortions were introduced in the beams. In an MCAO system,
several deformable mirrors (DMs) are used, optically conjugated to
various heights where the main turbulent layers are. The AO correction
is distributed among these DMs in order to enlarge the corrected
FOV.

This higher complexity of MCAO systems with respect to classical AO
has led to already several decades of research and developments on
tomographic reconstructors for MCAO. This paper gives a brief,
unfortunately non exhaustive, overview of the MCAO reconstructors
studied in the literature as well as the ones
implemented on sky so far. Section~\ref{sec:mcao-reconstr-probl}
reminds the fundamental equations and the main difficulties of the
MCAO reconstruction problem. Section~\ref{sec:inter-matr-based}
describes the reconstructors based on regularized inversions of the
interaction matrix of the MCAO system. Such reconstructors are the
ones implemented on existing MCAOs. Improved solutions based on
minimum-variance reconstruction are presented next in
Section~\ref{sec:minim-vari-reconstr}. In
Sect.~\ref{sec:three-step-reconstr}, the three-step reconstructors are
described together with their main limitation to cope with the effects
of lasers spot elongation. Finally,
Section~\ref{sec:kalman-filter-based} introduces the reconstructors
based on Kalman filters. They are expected to improve the AO correction
with respect to minimum-variance solutions, as soon as accurate model
approximation and temporal prediction of the turbulence could be
efficiently implemented.

\section{The MCAO reconstruction problem}
\label{sec:mcao-reconstr-probl}

This paper considers MCAO systems for which the measurements satisfy
an equation of the form
\begin{equation}
  \label{eq:MeasEq}
  \V{d} = \M{S} \M{\Gamma}^{g}\V{w} + \V{e}\,,
\end{equation}
where $\V{d}$ is the concatenated vector of measurements coming from
the various wavefront sensors. $\V{w}$ is a layered and discretized
representation of the wavefront distortions induced by the atmospheric
turbulence. $\M{\Gamma}^g$ is the linear operator of propagation of
the layered wavefront down to the pupil plane in the various guide
stars directions $g$. $\M{S}$ is the concatenated linear operator
representing the measurement devices, from pupil plane wavefronts to
local gradients. Finally, vector $\V{e}$ stands for the additive noise
on the data. Equation~(\ref{eq:MeasEq}) models the
measurements of locally averaged gradients in open-loop or pseudo
open-loop scheme. This can correspond to the use of Shack-Hartmann
wavefront sensors or Pyramid wavefront sensors in linear ranges. The
statistics of the turbulence-induced wavefronts $\V{w}$ and of the
noise $\V{e}$ is assumed to follow zero-mean Gaussian law of
respective covariances $\Cprior$ and $\Cerror$, such that:
\begin{equation}
  \label{eq:StatisticsWfAndNoise}
    \begin{bmatrix}
        \V{w} \\ \V{e}
    \end{bmatrix} \sim \mathcal{N}\left(\begin{bmatrix}
        \V{0} \\ \V{0}
    \end{bmatrix}, \begin{bmatrix}
        \Cprior & \V{0} \\
        \V{0} & \Cerror
    \end{bmatrix}\right)\,.
\end{equation}
The noise is uncorrelated with the turbulence-induced wavefronts.

The MCAO reconstruction problem consists in determining an estimate of
the turbulence-induced wavefront distortions $\hat{\V{w}}$. For this,
a linear estimator $\M{R}$ (a.k.a the reconstructor) is usually
applied to the data $\V{d}$, such that
\begin{equation}
  \label{eq:Reconst}
  \hat{\V{w}} = \M{R}\V{d}\,.
\end{equation}

Compared to classical AO reconstruction, we can already mention the following
specificities or requirements for an MCAO reconstructor:
\begin{enumerate}
\item the tomography, i.e. the reconstruction of wavefronts at
  various heights in $\hat{\V{w}}$
\item the goal to provide uniform AO correction over the science FOV instead of trying to zero the measurements
\item the null modes estimation when using laser guide stars (LGSs)
\item the accurate noise modeling and use of correlations to optimize the
  reconstruction in the presence of elongated spots with lasers
\end{enumerate}

\section{Interaction matrix-based reconstructors}
\label{sec:inter-matr-based}

The model in Eq.~(\ref{eq:MeasEq}) implicitly assumes a
parametrization of the wavefront distortions $\V{w}$. In all existing
AO systems, this parametrization is restricted to the degrees of
freedom of the DMs. In other words, $\V{w}$ represents the actuators
voltages on the DMs. In such approaches, the system is completely
characterized by the interaction matrix (IM) $\M{G}=\M{S}\M{\Gamma}^g$,
such that Eq.~(\ref{eq:MeasEq}) becomes
\begin{equation}
  \label{eq:MeasEqIM}
  \V{d} = \M{G} \V{w} + \V{e}\,.
\end{equation}

In practice, the IM $\M{G}$ is calibrated using Karhunen-Loeve modes
of the turbulence statistics projected on the DMs space
\cite{PetitEtAl2008a,EspositoEtAl2010a}. For this, the covariance
matrix $\Cprior$ is diagonalized and factorized in the form
\begin{equation}
  \label{eq:KLfactorization}
  \Cprior = \M{K} \M{K}\T\,,
\end{equation}
such that $\V{w} = \M{K} \V{u}$ represents a change of basis from the
rescaled Karhunen-Loeve modes $\M{u}$ to the corresponding actuators
voltages $\V{w}$. Using this change of variable, one can write again
the measurement equation (\ref{eq:MeasEqIM}) in the following way:
\begin{equation}
  \label{eq:MeasEqU}
  \V{d} = \M{G}\M{K} \V{u} + \V{e} = \M{G}_u \V{u} + \V{e}\,,
\end{equation}
with
\begin{equation}
  \label{eq:Statu}
   \begin{bmatrix}
        \V{u} \\ \V{e}
    \end{bmatrix} \sim \mathcal{N}\left(\begin{bmatrix}
        \V{0} \\ \V{0}
    \end{bmatrix}, \begin{bmatrix}
        \M{I} & \V{0} \\
        \V{0} & \Cerror
    \end{bmatrix}\right)\,.
\end{equation}
The linear reconstruction in this new modal space consists in
estimating
\begin{equation}
  \label{eq:ReconstU}
  \hat{\V{u}} = \M{R}_u \V{d} = \M{K}^{-1} \M{R} \V{d}\,.
\end{equation}

\subsection{Truncated-SVD reconstructor}
\label{sec:trunc-svd-reconstr}

The most common reconstructor implemented in existing MCAO systems is
based on the noise-weighted least-squares solution of
Eq.~(\ref{eq:MeasEqU}), that is
\begin{equation}
  \label{eq:TSVD-criterion}
  \hat{\V{u}}= \textrm{arg min}_u \norm{\V{d}-\M{G}_u\V{u}}^2_{\Cerror}\,.
\end{equation}
This best fit of the data is obtained using the singular value
decomposition (SVD) of the system matrix
$(\M{G}_u\T\Cerror^{-1}\M{G}_u)$ and computing its generalized inverse
$(\M{G}_u\T\Cerror^{-1}\M{G}_u)^{\dagger}$. The SVD reconstructor can
then be written as
\begin{equation}
  \label{eq:SVD-R}
  \M{R}_u^{\textrm{SVD}} = (\M{G}_u\T\Cerror^{-1}\M{G}_u)^{\dagger}\M{G}_u\T\Cerror^{-1}\,.
\end{equation}
In practice, the SVD reconstructor is not a suitable choice since it
usually includes singular values smaller than 1, indicating that the
singular modes are badly seen by the sensor device. The measurements
are more sensitive to noise than to the signal on these particular
modes. A trade-off is usually set between the best fit solution and
the mitigation of noise propagation in the reconstruction using a
truncated-SVD reconstructor
\begin{equation}
  \label{eq:TSVD-R}
  \M{R}_u^{\textrm{T-SVD}} = (\M{G}_u\T\Cerror^{-1}\M{G}_u)^{\dagger p}\M{G}_u\T\Cerror^{-1}\,,
\end{equation}
where the upperscript $p$ in the generalized inverse means that only
the $p$ largest singular-values have been kept in the inversion. The
other smaller singular values have been artificially set to zero. This
truncated-SVD solution only reconstructs the $p$ dominant modes of the
MCAO system.

Truncated-SVD reconstructors have been used in the natural guide
star-based MCAO systems MAD
\cite{RagazzoniEtAl2000a,RagazzoniDiolaiti2002a} (ESO-VLT) and
LINC-NIRVANA\cite{ZhangEtAl2012a} (LBT) as well as in solar MCAOs
\cite{SchmidtEtAl2016a}.

\subsection{Tikhonov reconstructor}
\label{sec:tikh-reconstr}

In other existing AO systems, the inversion of the system
$(\M{G}_u\T\Cerror^{-1}\M{G}_u)$ is done using Tikhonov
regularization. This is the case for the reconstruction part from the
laser guide stars data in Gemini Multiconjugate Adaptive Optics System
(GeMS) \cite{NeichelRigaut2014a} for instance.
The Tikhonov reconstructor then has the following expression
\begin{equation}
  \label{eq:Tik-R}
  \M{R}_u^{\textrm{Tik}} = (\M{G}_u\T\Cerror^{-1}\M{G}_u+\mu\M{I})^{-1}\M{G}_u\T\Cerror^{-1}\,,
\end{equation}
where there is no generalized inverse anymore. As a matter of fact,
the addition of the positive scalar $\mu$ on the diagonal of the symmetric
positive matrix of the system now allows it to become symmetric
positive definite and to have a uniquely defined inverse.

This reconstructor corresponds to modifying the best fit equation in
(\ref{eq:TSVD-criterion}) with a quadratic penalization term, thus
searching for
\begin{equation}
  \label{eq:Tik-criterion}
  \hat{\V{u}}= \textrm{arg min}_u \left[\norm{\V{d}-\M{G}_u\V{u}}^2_{\Cerror}+\mu \norm{\V{u}}^2 \right]\,.
\end{equation}

Since the prior covariance of $\V{u}$ is the identity matrix (see
Eq.~(\ref{eq:Statu})), the reconstructor in Eq.~(\ref{eq:Tik-R}) would
match the minimum-variance reconstructor when $\mu$ is set to 1. The
details about the derivation of the minimum-variance reconstructor are
presented below in Sect.~\ref{sec:minim-vari-reconstr}. In existing AO
systems, the reconstructor $\M{R}_u^{\textrm{Tik}}$ from
Eq.~(\ref{eq:Tik-R}) is applied on closed-loop data. The prior
covariance of the closed-loop (residuals) wavefront distortions differ
in general from the identity matrix. Therefore, even setting $\mu=1$,
this reconstructor cannot be a minimum-variance reconstructor when
applied on closed-loop data. In practice, the value of $\mu$ is tuned
to values smaller than 1 in order to mitigate the effect of using non
adequate statistics in the regularization.

It is worth mentioning that the Tikhonov reconstructor is used in GeMS
exclusively to reconstruct the part of the turbulence that is measured
by the lasers. This means that it is not used to estimate the null
modes associated to the tip-tilt anisoplanatism and tip-tilt
indetermination when using lasers. These six modes, namely the tip and
tilt for ground DM and the tip, tilt, focus and astigmatism for high
altitude DM, are obtained exclusively from the measurements on natural
guide stars (NGSs). A totally decoupled tomography from LGSs and from
NGSs is computed in GeMS. As mentioned in Neichel \textit{et al.}
\cite{NeichelRigaut2014a}, the reconstructed modes from the LGSs
measurements are
\begin{equation}
  \label{eq:GeMS-R-LGSs}
    \hat{\V{u}}^{L} = (\M{I}-\M{F}_u)(\M{G}^{L T}_u\Cerror^{L -1}\M{G}^L_u+\mu\M{I}+\beta \M{F}_u)^{-1}\M{G}_u^{L T}\Cerror^{L -1} (\M{I}-\M{F}_d)\,,
\end{equation}
where $(\M{I}-\M{F}_d)$ projects the LGSs measurements on an
orthogonal space to the null modes measurements, $\beta \M{F}_u$
penalizes the estimation of null modes and $(\M{I}-\M{F}_u)$ projects
the estimated wavefronts orthogonally to the null modes space. This
prevents any contribution to null modes to be included in
$\hat{\V{u}^L}$. The use of the upperscript $^{L}$ in
Eq.~(\ref{eq:GeMS-R-LGSs}) stands for the restrictions of $\M{G}_u$
and $\Cerror$ to their projection on the space orthogonal to the null
modes and to the measurements of the null modes. In GeMS, the null
modes are reconstructed only from the NGSs measurements, optimizing
the null modes correction by all DMs from an estimate of the tip-tilt
residuals in 9 directions among the FOV. This is interesting since it
is closer to a minimum-variance approach for the null modes estimation
and correction than what is done for the LGSs part. This can have a
positive contribution to the good uniformity of GeMS correction
among its FOV. A deeper analysis of this ad-hoc decoupling and the
different reconstruction schemes would still be necessary to draw
clear conclusions on this aspect.

\subsection{Limitations of IM-based reconstructors}
\label{sec:mcaos-as-extensions}

\begin{figure} [ht]
   \begin{center}
     \begin{minipage}[c]{0.44\linewidth}
   \begin{tabular}{c} 
   \includegraphics[height=10cm]{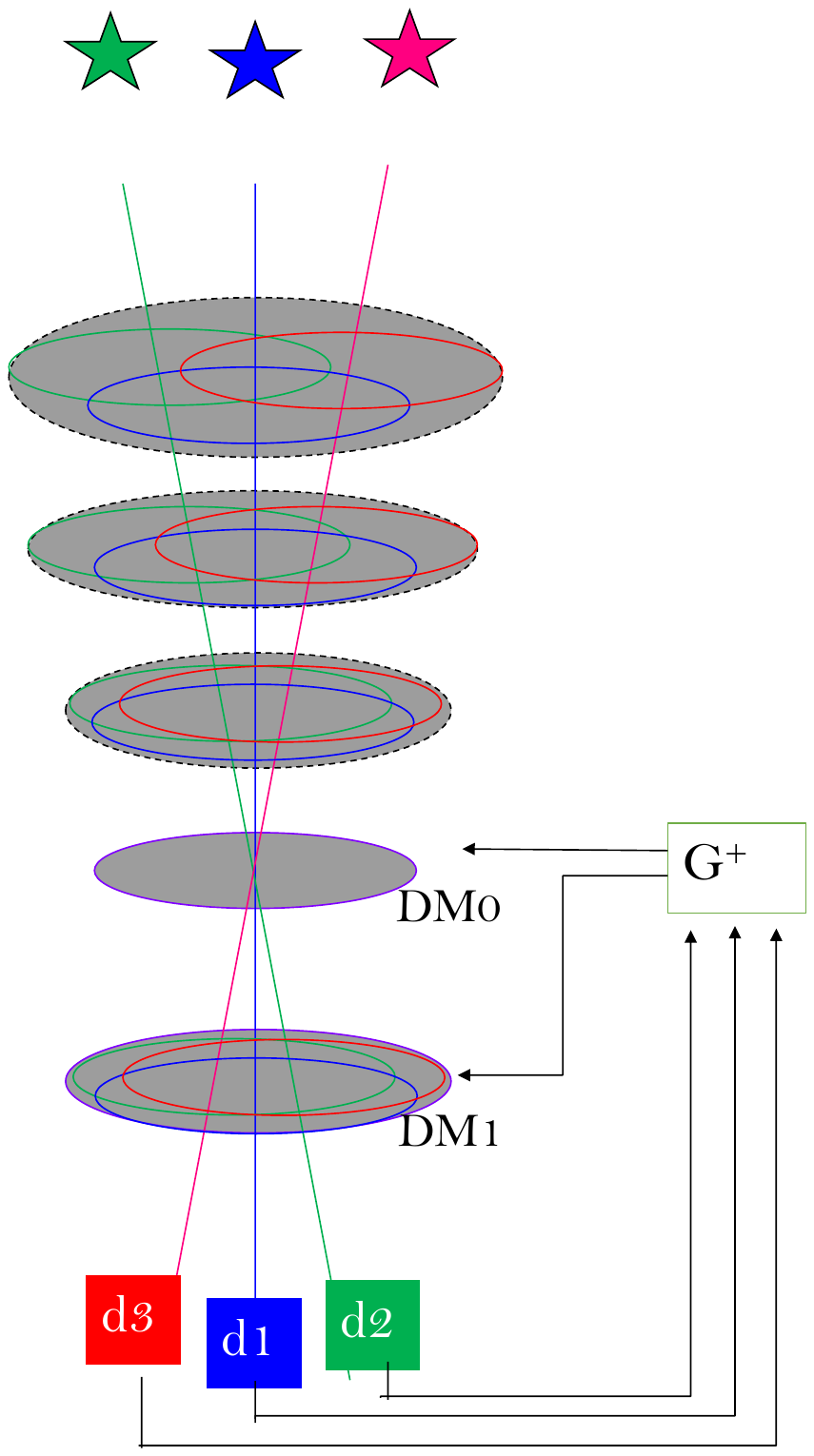}
   \end{tabular}
     \end{minipage}
\vline
     \begin{minipage}[c]{0.52\linewidth}
   \begin{tabular}{c} 
   \includegraphics[height=11cm]{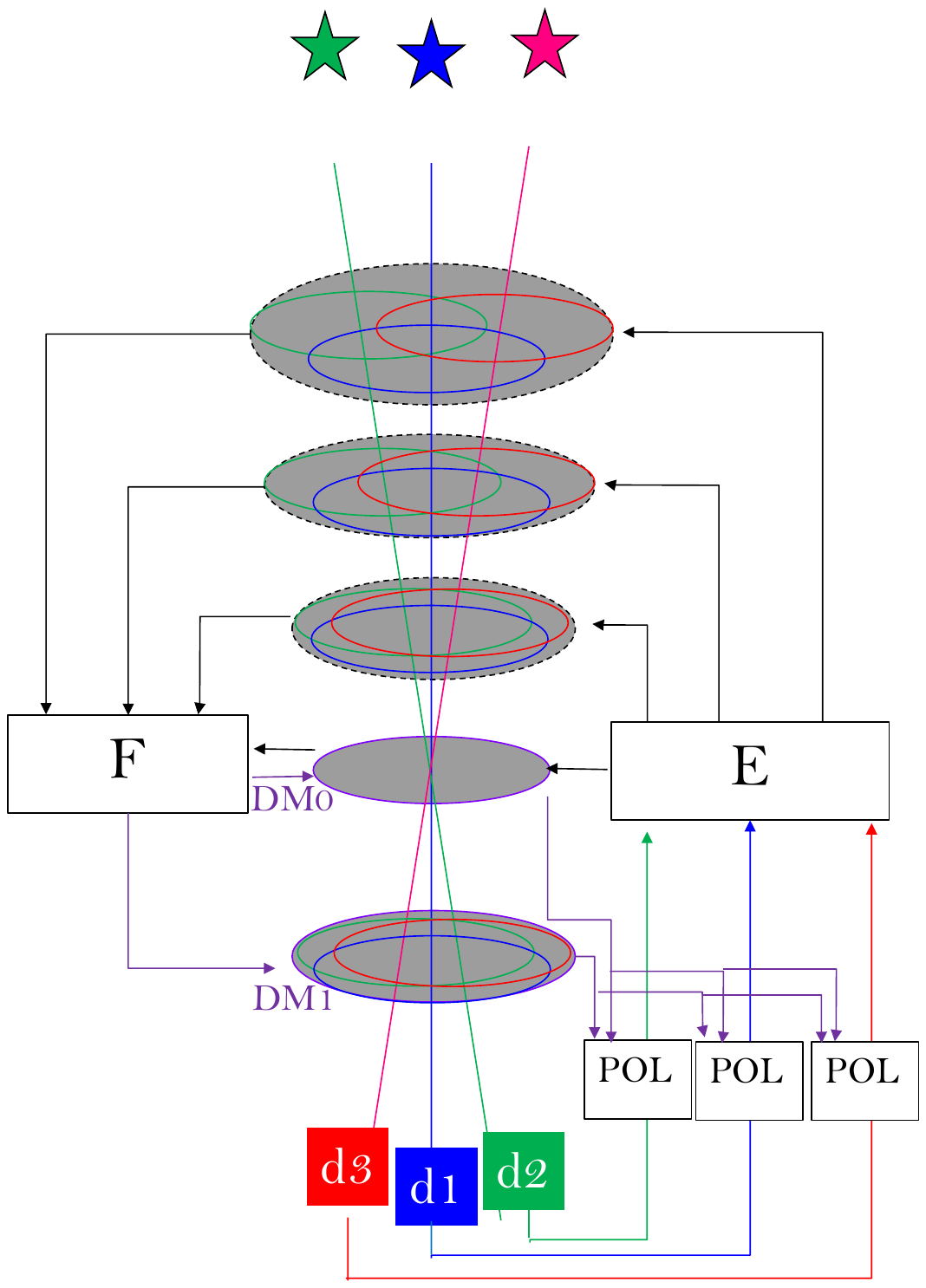}
   \end{tabular}
     \end{minipage}
  \end{center}
   \caption[IM-basedR-and-MinVar-schemes]
   { \label{fig:IM-basedR-and-MinVar-schemes} Left: Summarized scheme
     for an IM-based reconstruction in star-oriented MCAO. The
     wavefront sensors data are processed altogether by a regularized
     inverse of the IM of the system in order to deliver directly the
     incremental commands to be sent to the DMs. For clarity reasons,
     the integrator-like control implemented before sending the
     control vectors to the DMs is not represented in the
     diagram. Right: Summarized scheme for a minimum-variance
     reconstructor in star-oriented MCAO. First, estimates of pseudo
     open-loop data are computed. Applying $\M{E}$ consists in solving
     the tomographic reconstruction problem from slopes space to
     wavefront space. The application of $\M{F}$ corresponds to the projection
     step from the reconstructed layers to the DM controls space.}
   \end{figure}

We are currently in the early era of MCAO systems when the
reconstructors implemented on-sky are the two IM-based reconstructors
described above. The main principles behind these approaches for MCAO
is summarized on the left part of
Fig.~\ref{fig:IM-basedR-and-MinVar-schemes}, where the wavefront
sensors data are processed altogether by a regularized inverse of the
IM of the system in order to deliver directly the control vectors to
be sent to the DMs. Note that in the present paper, the configuration
of layer-oriented MCAO \cite{RagazzoniDiolaiti2002a} is not
specifically considered, although the reconstructors involved in
existing layer-oriented systems (e.g. MAD, LINC-NIRVANA) are based on
similar principles with truncated-SVD reconstructors for each layer
independently \cite{ZhangEtAl2012a}.
The main reason for this is
understood to be because they both correspond to what had been
implemented previously in single-conjugate AO systems on large
telescopes \cite{DamEtAl2004a,EspositoEtAl2010a,SauvageEtAl2016a}, but
extended to the MCAO configuration. The IM-based approach also has the
practical advantage that there is no need to accurately model the
system geometry, since it is implicitly included in the
calibrated matrices. The T-SVD reconstructor allows to correct up to
$p$ modes. One can modify ad-hoc the value of $p$ according to the design
of the system or its observing conditions. In a similar way, the
tuning of the scalar $\mu$ in Tikhonov reconstructor allows to vary
the regularization depending on the signal-to-noise ratio. The
Tikhonov approach had shown interesting results in classical AO at the
Keck Observatory \cite{DamEtAl2004a} to reduce the reconstruction of
local waffles compared to zonal T-SVD previously implemented on this
system.

However, it is important to be aware that these reconstructors are
suboptimal MCAO reconstructors. The optimal reconstruction is a
minimum-variance reconstruction as described further in
Sect.~\ref{sec:minim-vari-reconstr}, where the prior statistics of the
incoming signal is used for regularization instead of an ad-hoc
regularization (truncation or Tikhonov here).

In addition, the IM-based reconstructors are implemented in existing
MCAO systems with integrator-like controls on residuals
wavefronts. Such controls are defined aiming to zero the closed-loop
measurements. This goal is different from providing a uniform AO
correction among the science FOV. In other words, the AO correction is
optimized for the guide stars, but not for other possible science
targets. Using those reconstructors for solar or NGS-based MCAOs, the
uniformity of the AO correction relies on the use of many guide stars
distributed regularly among the FOV. Separation between guide stars
must be carefully chosen to avoid the generalized anisoplanatism to
reduce the correction quality in between the guide stars
directions. For laser-based MCAO, the AO correction is no longer
optimized for science since it is optimized in directions of stars
located at a finite range.

Last, in IM-based reconstructors, the estimation is always made at the
heights of the optical conjugation of the DMs. These heights are
usually fixed in the design of the MCAO systems, although the
atmospheric turbulent layers are located at heights that vary with
time, at minute and day scales, and with the observing elevation
angle. The few fixed DMs heights induce an intrinsic generalized
fitting error; a characteristic of the MCAO design itself at a given
observing site. The only way to reduce the generalized fitting error
is to add degrees of freedom on the correction, that is to say usually
adding more high-altitude DMs. Another error source appears in
IM-based reconstruction because the tomographic reconstruction is
restricted to the DMs heights \cite{FuscoEtAl2001a}. This tomographic
error could be reduced by reconstructing the disturbances over a
larger number of layers located closer to where the strongest
turbulent layers actually are. Next, using the reconstructed layers,
one can derive the best control vector to send to the DMs to provide a
uniform AO correction over the FOV. This final projection computation
\cite{FuscoEtAl2001a}, also known as fitting step
\cite{YangVogel2006b}, is detailed in the reconstructors described
below.

\section{Minimum-variance reconstructors}
\label{sec:minim-vari-reconstr}

Contrary to IM-based reconstructors, the minimum-variance
reconstructors are derived from the optimization of a criterion
related to the quality of the AO correction among a scientific
FOV. Such MCAO criterion is usually formulated in the following way
\cite{Ellerbroek2002a}
\begin{equation}
  \label{eq:MCAO-crit}
  \epsilon^2 = \avg{\V{\phi}^{sT} \M{W} \V{\phi}^s  }_{w,e}\,,
\end{equation}
where ${\phi}^{s}$ stands for the pupil plane resulting phases along a
set of directions discretizing the scientific FOV, $\M{W}$ is a linear
operator removing the piston contribution to the resulting phases and
applying a weighting according to the importance of each direction -
usually a uniform weighting is expected for MCAO. Notation $\avg{.}_{w,e}$
stands for the expected value with respect to turbulence and noise
statistics. The resulting phases $\phi^s$ in science directions can be
expressed as
\begin{equation}
  \label{eq:SciencePhases}
  \V{\phi}^s = \M{\Gamma}^s \V{w} - \M{M}_a^s \V{a}\,,
\end{equation}
where $\M{\Gamma}^s$ is the operator propagating the layered
wavefronts in the science direction towards the pupil plane, $\V{a}$
stands for the control vector of the DMs and the operator $\M{M}^s_a$
models the relationship between control vector and resulting
correction phases in pupil plane in the same science directions. The layered wavefronts $\V{w}$ are
modeled in a different space (usually of greater dimension) than the
one of the control vector $\V{a}$.

The goal is to estimate the control
vector $\V{a}=\M{R}\V{d}$, linearly from the data, minimizing the AO
criterion $\epsilon^2$ of Eq.~(\ref{eq:MCAO-crit}), i.e.
\begin{equation}
  \label{eq:Min-Var}
  \hat{\V{a}}^{*} = \textrm{arg min}_{a}  \epsilon^2 = \textrm{arg min}_{a} \avg{(\M{\Gamma}^s \V{w} - \M{M}_a^s \M{R}\V{d})\T \M{W} (\M{\Gamma}^s \V{w} - \M{M}_a^s \M{R}\V{d})}_{w,e}\,.
\end{equation}
This leads to a more general formulation of the MCAO reconstruction
problem than what is proposed in
Eq.~(\ref{eq:Reconst}). Distinguishing wavefronts $\V{w}$ from
commands $\V{a}$ allows a more accurate modeling of the system, its
disturbances and its control.

\subsection{General expression of the Reconstructor}
\label{sec:reconstr-struct}

The optimization of the above criterion leads to a reconstructor
$\M{R}=\M{F}\M{E}$, that can be decomposed in 2 successive operators
\begin{itemize}
\item $\M{E}$, the maximum a posteriori estimator of the layered
  wavefront $\hat{\V{w}}$;
\item $\M{F}$, the optimal projector of the layered wavefront
  estimates $\hat{\V{w}}$ on the control space of the DMs.
\end{itemize}
The first operator $\M{E}$ is the maximum a posteriori estimator of
the layered wavefront according to the model in Eq.~(\ref{eq:MeasEq}),
which has the following equivalent expressions
\cite{Wallner1983a,FuscoEtAl2001a,Ellerbroek2002a,ThiebautTallon2010a}
\begin{eqnarray}
  \label{eq:MinVarWF-E}
  \M{E} & = &\avg{\V{w} \V{d}\T} \, \avg{\V{d} \V{d}\T}^{-1} \\
 & = &\Cprior \M{\Gamma}^{gT} \M{S}\T (\M{S}\M{\Gamma}^{g} \Cprior \M{\Gamma}^{gT}\M{S}\T + \Cerror)^{-1} \label{eq: MinVarWF-E-slopes}\\
& = &(\M{\Gamma}^{gT}\M{S}\T \Cerror^{-1} \M{S}\M{\Gamma}^{g} + \Cprior^{-1})^{-1} \M{\Gamma}^{gT}\M{S}\T \Cerror^{-1} \label{eq:MinVarWF-E-wf}\,.
\end{eqnarray}
The optimal projector $\M{F}$ has the following expression
\begin{equation}
  \label{eq:OptimProj-F}
  \M{F} = (\M{M}_a^{sT} \M{W}\M{M}_a^{s} )^{\dagger} \M{M}_a^{sT} \M{W}\M{\Gamma}^s_w\,.
\end{equation}
In general, the inversion involved in the projector is not uniquely
defined (symbolized by the upperscript $^{\dagger}$), since modifying
the control to produce piston over the mirrors for instance does not
modify the correction criterion value. In practice, one would like to
minimize the control voltages to be applied, which leads to a Tikhonov
regularization of the projector formulated as
\cite{Ellerbroek2002a,YangVogel2006b}
\begin{equation}
  \label{eq:OptimProj-F-penalized}
  \M{F} = (\M{M}_a^{sT} \M{W}\M{M}_a^{s} +\alpha \M{I})^{-1} \M{M}_a^{sT} \M{W}\M{\Gamma}^s_w\,.
\end{equation}

\subsection{Pseudo open-loop data}
\label{sec:pseudo-open-loop}

In practice, the MCAO systems work in closed-loop. The measurement
equation in (\ref{eq:MeasEq}) is thus not an exact modeling of the
system. The residual, closed-loop, measurements are in fact
described by the following model
\begin{equation}
  \label{eq:MeasEq-CL}
  \V{d}^{\textrm{CL}}(k) = \M{S} \M{\Gamma}^{g} \V{w}(k-\tau) - \M{G} \V{a}(k-\tau) + \V{e}(k)\,,
\end{equation}
where the index $k$ indicates the discrete time step associated to a
set of measurements. The measurement process induces some delay $\tau$
(e.g. exposure time, reading and computations) which is usually close
to 2 frames in existing systems. Note that Eq.~(\ref{eq:MeasEq-CL})
makes the interaction matrix of the system $\M{G}$ appear, although it
was not used above for the derivation of the minimum-variance
reconstructor.

In order to obtain a minimum-variance reconstruction, it is required
to transform the closed-loop data $\V{d}^{\textrm{CL}}$ into an
estimate of pseudo open-loop data $\V{d}^{\textrm{POL}}$. Such
estimate aims to fit the model of Eq.~(\ref{eq:MeasEq}) used as a
basis for the derivation of the minimum-variance reconstruction
operators $\M{E}$ and $\M{F}$. At every frame, pseudo open-loop
measurements $\V{d}^{\textrm{POL}}$ are thus computed as
\begin{equation}
  \label{eq:POL}
  \V{d}^{\textrm{POL}}(k) = \V{d}^{\textrm{CL}}(k) + \M{G} \V{a}(k-\tau)\,.
\end{equation}
The minimum-variance reconstruction is then computed as
\begin{equation}
  \label{eq:MinVar-on-POL-data}
  \hat{\V{a}}(k) = \M{F}\M{E}\V{d}^{\textrm{POL}}(k)\,.
\end{equation}

The use of pseudo open-loop data is usually combined with an ad-hoc
control, kwown in AO as pseudo open-loop control scheme
\cite{Gilles2005a}. It allows to include an
integrator-like feedback.

\subsection{Implementation methods for the wavefront estimator}
\label{sec:impl-meth}

A summary of the minimum-variance reconstruction steps together with
the computation of the pseudo open-loop measurements is illustrated on
the right part of Fig.~\ref{fig:IM-basedR-and-MinVar-schemes}. The
pseudo open-loop computation is an additional computation requirement
compared to IM-based reconstruction. Nevertheless, it is key to note
that pseudo open-loop computations from Eq.(\ref{eq:POL}) are based on
previous control vector $\V{a}(k-\tau)$, thus providing two frames of
time to precompute the additional term $\M{G}\V{a}(k-\tau)$ before the
newcoming data $\V{d}^{\textrm{CL}}(k)$ become available.

The increased complexity of minimum-variance reconstruction through
steps $\M{E}$ and $\M{F}$ compared to the IM-based reconstruction (see
the two sides of Fig.~\ref{fig:IM-basedR-and-MinVar-schemes}) has been
a matter of concern and a subject of research during the last two
decades \cite{Ellerbroek2002a}. The implementation of these two steps
can in practice be formulated as a single matrix-vector
multiplication. Until recently, it was not guaranteed that the
developments of the real-time computers of the GSMTs could handle the
multiplication of such matrix by a vector at the scale of an MCAO
system for the GSMTs. The recent advances in GPUs and multiple CPUs
architectures may however provide the required technology to implement
matrix-vector multiplication for some future MCAO systems. In
particular, the baseline for the MCAO system of the Thirty Meter
Telescope (TMT), NFIRAOS, is to use a matrix-vector multiplication to
apply $\M{R}=\M{F}\M{E}$ to the pseudo open-loop measurements
\cite{WangEllerbroek2012a}.

The challenge also consists in precomputing the matrices $\M{F}$ and
$\M{E}$ fast enough, since they need to be updated when the system
geometry evolves at times scales shorter than a minute (e.g. field
rotations, variations of altitude of the main atmospheric
layers). Many algorithms have been studied during the past two decades
to try to accelerate the computation of these operators.

In 2002, B. Ellerbroek presented a direct computation of the maximum a
posteriori estimator and the projection operator using sparse matrix
techniques \cite{Ellerbroek2002a}. This work reduced the complexity of
the precomputation of matrix $\M{E}$ to order $N^{3/2}$ operations
instead of $N^3$ for classical inversion methods, with $N$ the number
of degrees of freedom of the AO system. It uses the fact that most of
the numerical models involved in an MCAO system have sparsity
properties. As a matter of fact, most of the devices (shack-Hartmann
sensors $\M{S}$, mirror actuators $\M{M}$, interaction matrix $\M{G}$)
and operators (propagators $\M{\Gamma}$) only act locally. The main
difficulty is to obtain a sparse approximation of the
turbulent wavefront statistics, $\Cprior$. For this, the covariance is
modeled as a discrete Laplacian operator \cite{Ellerbroek2002a}. This
approach is the baseline for the implementation of the computation of
$\M{E}$ for NFIRAOS on the TMT \cite{WangEllerbroek2012a}.

As possible alternatives to the computation of the inverse matrix in
Eq.~(\ref{eq:MinVarWF-E-wf}), many iterative methods have been studied
since 2000 to estimate the layered wavefront
$\hat{\V{w}}=\M{E}\V{d}^{\textrm{POL}}$. All these methods consist in iteratively
solving the same linear system
\begin{equation}
  \label{eq:LinSys-iterative}
  (\M{\Gamma}^{gT}\M{S}\T \Cerror^{-1} \M{S}\M{\Gamma}^{g} + \Cprior^{-1}) \V{w} = \M{\Gamma}^{gT}\M{S}\T \Cerror^{-1} \V{d}^{\textrm{POL}}\,.
\end{equation}
They mainly differ by the way they approximate the inverse of the
covariance matrix $\Cprior$. This limiting operation finally defines
the order of the entire method. To be considered as an interesting
candidate for implementation at a GSMT scale, an iterative algorithm
for MCAO reconstruction is expected to scale at most as
$\mathcal{O}(N^{3/2})$, $\mathcal{O}(N \log N)$ or, better,
$\mathcal{O}(N)$ in terms of number of operations.

They essentially use the conjugate gradients (CG) algorithm or the
preconditioned conjugate gradients (PCG)
\cite{GillesEllerbroek2003a}. The minimum-variance tomographic
reconstruction as represented by Eq.~(\ref{eq:LinSys-iterative}) is
ill-conditioned. The convergence of the CG get slower when the number
of degrees of freedom of the system increases. For instance, 30
iterations of CG are at least necessary for NFIRAOS size and even more
for an MCAO system on the ELT. To meet the low-latency requirements of
AO correction, reducing the number of iterations is key. Large efforts
have been put in the past decades in determining efficient
preconditioners for large scales AO systems. The chosen approximation
for the covariance matrix $\Cprior$ also usually affects the
convergence of the algorithm when a preconditioner can be derived out
of it. A list of fast PCG-based algorithms for minimum-variance
estimation of the layered wavefronts is
\begin{itemize}
\item Block symmetric Gauss-Seidel(BSGS) or Multigrid (MG) PCG \cite{GillesEllerbroek2003a,GillesVogel2002b}
\item Fourier-Domain PCG (FD-PCG) \cite{YangVogel2006a}
\item Fractal iterative method (FRiM) \cite{ThiebautTallon2010a,TallonEtAl2010a}
\item Finite-element Wavelet hybrid algorithm (FEWHA) \cite{HelinYudytskiy2013a}.
\end{itemize}

In general terms, the common advantages of all these iterative methods are
\begin{itemize}
\item to avoid the direct inversion of huge matrices
\item to benefit from the sparsity of most of the involved AO operators
\item to have the flexibility to adapt the direct model on the fly
\item to optimize the tomography taking into account noise correlations when using lasers with elongated spots \cite{TallonTallon-Bosc2008b,GillesWang2010a}
\item to allow optimal combination of the LGS and NGS measurements when using minimum-variance tomography \cite{GillesEllerbroek2008a}
\item to converge in few iterations using warm start technique
  \cite{LessardWest2008a}
\end{itemize}
There exist no published study comparing all these algorithms together
in a same MCAO configuration. The various consorcia involved in the
design of GSMTs MCAO systems have however performed partial
comparisons between those, showing that in general they provide very
similar quality of AO correction
\cite{LessardWest2008a,GillesEllerbroek2003a,WangEllerbroek2012a,RamlauEtAl2014a}. In
addition, numerical simulations have been published for several MCAO
configurations at GSMT scale (TMT and ELT) showing how the
minimum-variance estimators outperform the IM-based reconstructors
\cite{FuscoEtAl2001a,HelinYudytskiy2013a}.

\subsection{Implementation methods for the projection step}
\label{sec:import-proj-step}

Following similar reasoning as for the wavefront estimation step, the
projection step defined by Eqs.~(\ref{eq:OptimProj-F}) or
(\ref{eq:OptimProj-F-penalized}) can be implemented using sparse
matrix techniques \cite{Ellerbroek2002a} or iterative methods
(e.g. PCG). For this, one takes advantages again from the fact that
the actuators have local influence on the DMs shape, leading to a
sparse operator $\M{M}_a^{s}$. The linear system to solve
\begin{equation}
  \label{eq:LinSys-Fitting}
  (\M{M}_a^{sT} \M{W}\M{M}_a^{s} +\alpha \M{I}) \V{a} = \M{M}_a^{sT} \M{W}\M{\Gamma}^s_w
\end{equation}
is better conditioned than the one of the tomgraphic wavefront
reconstruction allowing much better convergence rate. A few iterations
are needed in general using only a Jacobi preconditioner. A
Fourier-Domain preconditioner \cite{YangVogel2006b} has also been
proposed to optimize the performance in only 1 iteration.

\begin{figure} [ht]
   \begin{center}
   \begin{tabular}{c} 
   \includegraphics[height=7cm]{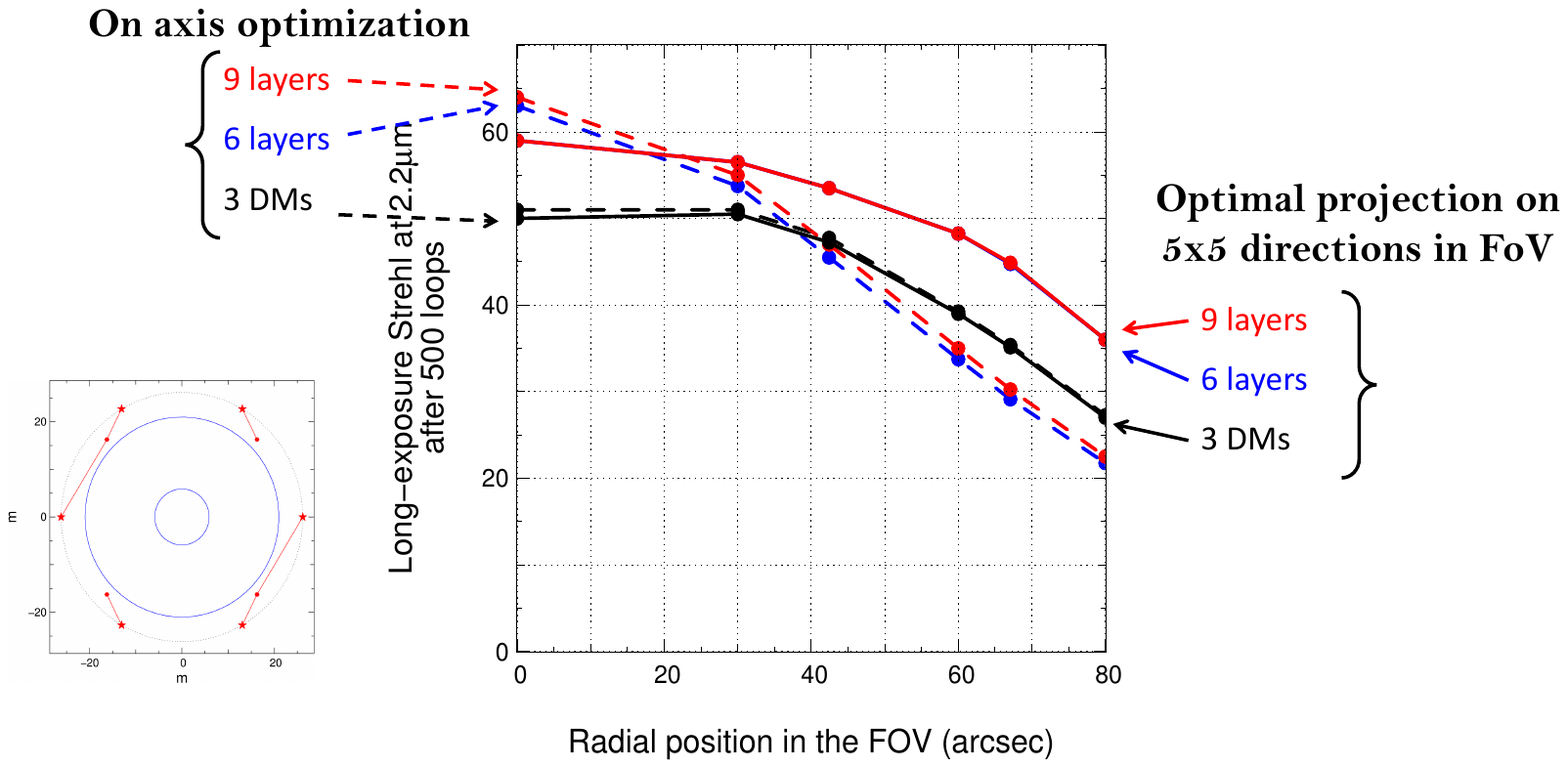}
   \end{tabular}
  \end{center}
   \caption[MCAO-projection]
   { \label{fig:MCAO-projection} End-to-end simulations results for
     ELT MCAO system MAORY using Octopus (ESO end-to-end AO simulator)
     and FRiM for the minimum-variance reconstruction and control. The
     left bottom frame illustrates the positions of the LGSs (red
     stars) and laser launch telescopes (red circles) in the
     meta-pupil. The radial distribution of the Strehl ratio (at
     2.2$\mu$m) is presented depending on the number of reconstructed
     layers (3 layers where the DMs are conjugated in red - 6 layers
     in blue - 9 layers in black) and on the projection step on the
     DMs space. Dashed lines correspond to optimization of the
     projection for on-axis direction only. Solid lines correspond to
     optimization of the correction simultaneously in $5\times 5$
     directions among the FOV. Note that in the case of only 3
     reconstructed layers at the altitude of conjugation of the DMs,
     the projection step has no effect. This 3-DM performance is similar to
     the one that could achieve at best IM-based reconstructors if they would
     be implemented with POLC control.}
   \end{figure}

The importance of this fitting step has been evidenced in several MCAO
studies. IM-based reconstructors presented in
Sect.~\ref{sec:inter-matr-based} usually provide relatively uniform
Strehl ratios among the FOV, if the guide stars asterism is suitably
chosen. Greater Strehl ratios could however be achieved when using
minimum-variance wavefront tomographic reconstruction on more layers
than the number of existing DMs. Obtaining uniformly distributed and
higher Strehl ratios among the FOV is however only possible when
combining the minimum-variance estimator together with the optimized
projection on DMs. This typical behavior is illustrated in
Fig.~\ref{fig:MCAO-projection} using end-to-end simulations of MAORY
configuration, the MCAO system for the ELT. The increased number of
the reconstructed layers together with the optimized projection step
allow to increase the Strehl ratio by 20\% among the FOV.

\section{Three-step reconstructors}
\label{sec:three-step-reconstr}

\begin{figure} [ht]
   \begin{center}
     \begin{minipage}[c]{0.48\linewidth}
   \begin{tabular}{c} 
   \includegraphics[height=10cm]{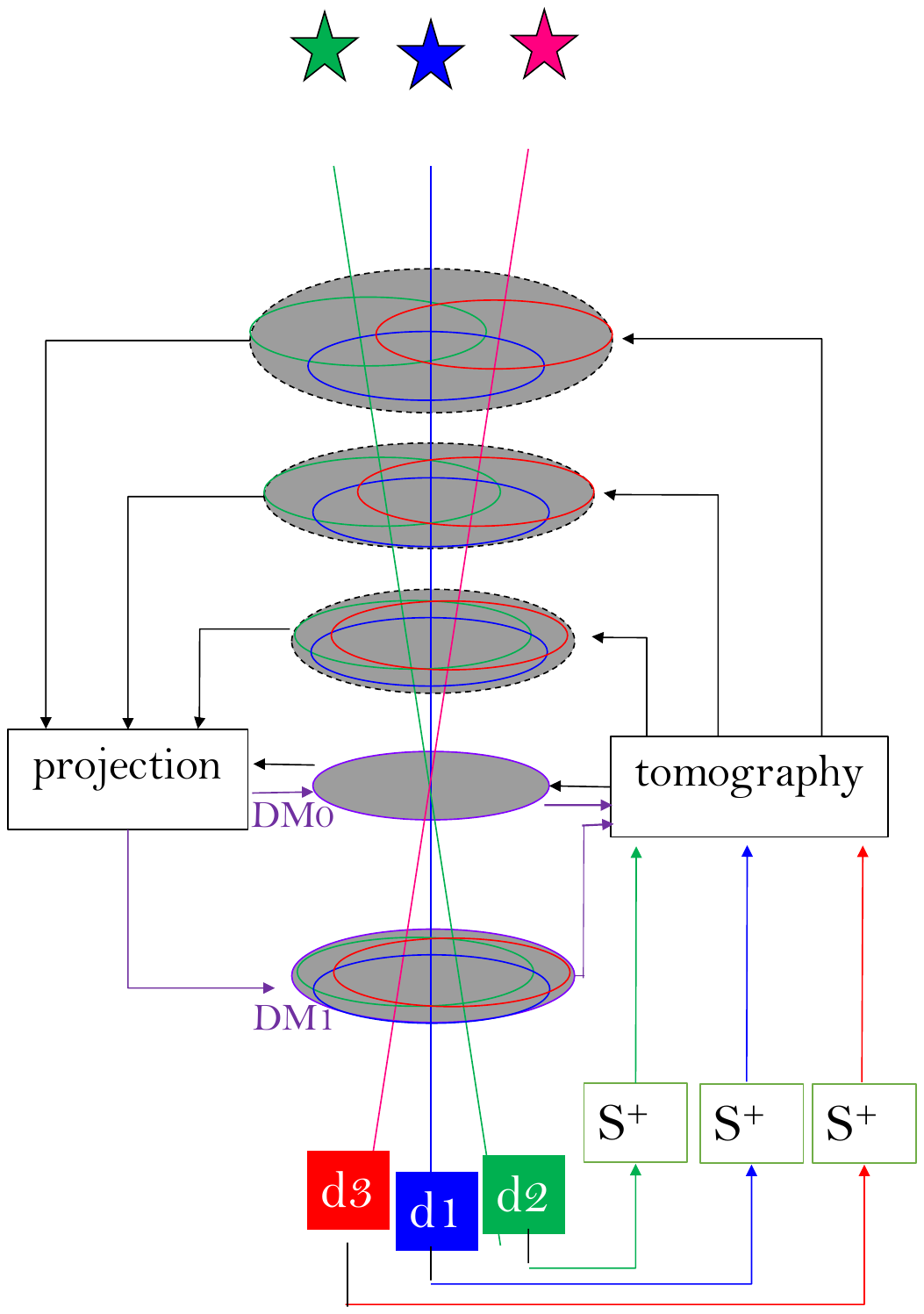}
   \end{tabular}
     \end{minipage}
\vline
     \begin{minipage}[c]{0.48\linewidth}
   \begin{tabular}{c} 
   \includegraphics[height=10cm]{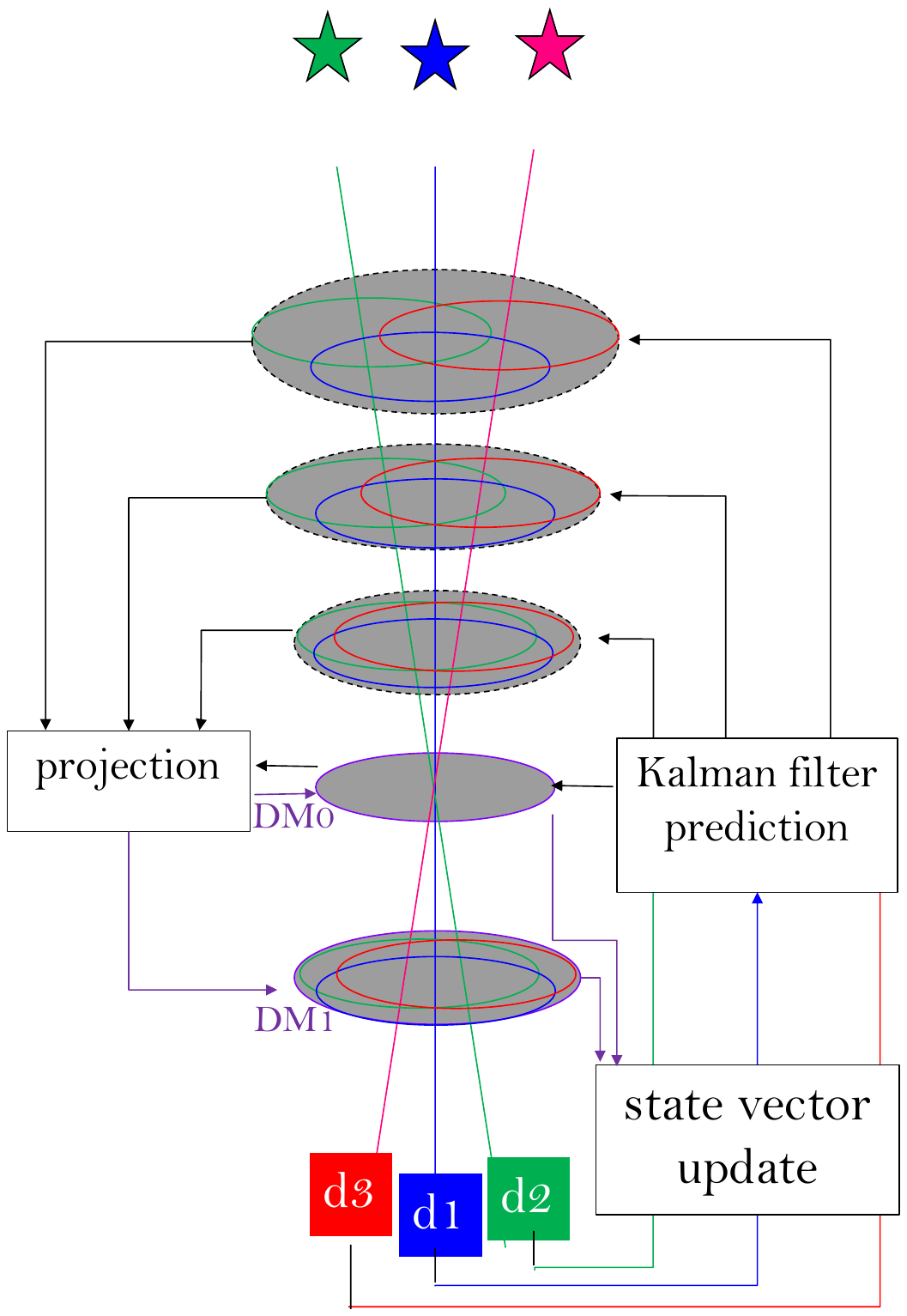}
   \end{tabular}
     \end{minipage}
  \end{center}
   \caption[3stepR-and-KalmanR-schemes]
   { \label{fig:3stepR-and-KalmanR-schemes} Left: Summarized scheme
     for an 3-step reconstructor in star-oriented MCAO. Step 1
     consists in computing the pupil-plane wavefront distortions
     independently for each guide star direction. Step 2 consists in
     solving the tomography problem in wavefront space using the
     reconstructed wavefront from all guiding directions. Step 3
     corresponds to the fitting or projection step from the
     reconstructed layers to the DM controls space. Right: Summarized
     scheme for a Kalman filter based reconstruction in MCAO. }
 \end{figure}

During the last decade and the design studies for the ELT, a new kind
of MCAO reconstructors has been proposed based on a 3-step
architecture \cite{RosensteinerRamlau2013a,RamlauEtAl2014a} as
represented on the left part of
Fig.~\ref{fig:3stepR-and-KalmanR-schemes}. The 3 steps are:
\begin{enumerate}
\item pupil-plane wavefront reconstructions from closed-loop data in each guide star direction independently,
\item wavefront tomography from pseudo open-loop pupil-plane
  wavefronts to layers,
\item projection from layers to DMs control space.
\end{enumerate}
From the comparison of the left schemes of
Figs.~\ref{fig:IM-basedR-and-MinVar-schemes} and
\ref{fig:3stepR-and-KalmanR-schemes}, it already appears that a
3-step reconstructor intends to correct for a major weakness of the
IM-based reconstructors. It explicitly includes a projection step to
optimize the AO correction among the FOV (Step 3). In that sense, the
3-step reconstructors look more similar to minimum-variance approach
(see right of Fig.~\ref{fig:IM-basedR-and-MinVar-schemes}). On the
contrary, the separation of step 1 and step 2 in the 3-step
reconstructors rather comes from latency considerations
\cite{RamlauEtAl2014a} than from AO correction optimization. 

\subsection{Implementations of each step}
\label{sec:impl-each-step}

Step 1 tackles the closed-loop wavefront reconstruction from
each guide star independently. This means that for the $i$-th guide star, the associated measurement equation
\begin{equation}
  \label{eq:MeasEq-GSg}
  \V{d}^{gi} = \M{S}\V{w}^{gi} + \V{e}^{gi}
\end{equation}
is considered. The pupil-plane resulting wavefront $\V{w}^{gi}$ for $i$-th
direction is reconstructed using a pseudo-inversion of
Eq.~(\ref{eq:MeasEq-GSg}), illustrated in
left part of Fig.~\ref{fig:3stepR-and-KalmanR-schemes} by $\M{S}^{+}$. Compared
to Eq.~(\ref{eq:MeasEq}), there is no propagation operator
$\M{\Gamma}^{g}$ in this step.

In the studies of the 3-step reconstructors, step 1 is usually done with
the fast order-$n$ algorithm named CuReD \cite{Rosensteiner2012a}. As
a matter of fact, if one wants to implement a sequence of 3 computing
steps for the reconstruction and compete with the latency of a single
matrix-vector multiplication (see left of
Fig.~\ref{fig:IM-basedR-and-MinVar-schemes} for instance), each of the
3 steps must involve very few computations. The CuReD algorithm is
highly pipelinable and parallelizable making it a suitable candidate
for such 3-step approach in terms of latency.

Step 2 corresponds to the wavefronts tomography, from pupil-plane
estimated wavefronts coming from various directions to wavefront
distortions on layers distributed in altitude. The estimated residuals
wavefronts are gathered in vector
$\hat{\V{w}}^g=[\hat{\V{w}}^{g1 T}, \hat{\V{w}}^{g2 T}, ...,
\hat{\V{w}}^{gn T}]\T$. Pseudo open-loop pupil-plane wavefronts
are computed removing the past contribution of the DMs to this estimated
wavefronts, such that
\begin{equation}
  \label{eq:POL-pupil-plane-wfs}
  \hat{\V{w}}^{g \textrm{POL}} = \hat{\V{w}}^g + \M{\Gamma}^{g} \V{a}\,.
\end{equation}

The wavefront distortions $\V{w}$ discretized among various layers
(i.e. a layered wavefront) are then estimated solving the linear system
\begin{equation}
  \label{eq:WF-Tomography}
  \M{\Gamma} \V{w} = \V{w}^{g \textrm{POL}}\,,
\end{equation}
where $\M{\Gamma}$ is a concatenation of the propagation operators
from the layers to the pupil-plane in all guide stars directions. This
tomography problem is ill-posed and requires regularization to obtain
a stable solution. Two iterative algorithms have mainly been studied in the
literature to implement this tomographic step:
\begin{enumerate}
\item a Kaczmarz method,
\item a Gradient-based method.
\end{enumerate}
A few characteristics of these methods are provided here in order to
understand how they differ one from another
\cite{RamlauEtAl2014a}. The Kaczmarz method is regularized only by
stopping early the iterations . It is thus difficult to draw any
general conclusions on the stability of this approach. The
gradient-based method includes a regularization term based on the a
priori statistics of the turbulent wavefronts. It also includes a
model of errors (propagated noise) to weight and whiten the incoming
pupil-plane wavefront estimates in the tomography.

An original feature of the Kaczmarz method is that it processes the
tomography in the various directions sequentially. The gradient-based
method on the contrary process the tomography from estimated
pupil-plane wavefront in all directions at the same time.

Step 3 corresponds to the projection of the reconstructed layered
wavefront on the control space of the DMs. This step is the same as
the one presented in Sect.~\ref{sec:import-proj-step} for the
minimum-variance reconstructors.

\subsection{Limitations of the 3-step reconstructors}
\label{sec:limitations-3-step}

The major limitation of the 3-step approach is that the information of
the noise correlations in the measurements is not taken into account
in processing the first step. As a consequence, suboptimal wavefront
reconstruction is obtained and all the subsequent steps will suffer
from this early degraded reconstruction. Although noise correlations
are usually negligible in a single-conjugate AO sytem, they are
crucial in tomographic reconstruction with LGSs for the GSMTs. The
laser elongated spots induce high measurement uncertainty in the
elongation direction and the use of an accurate model of the noise
covariance (with a minimum-variance reconstructor as in
Sect.~\ref{sec:minim-vari-reconstr}) has been demonstrated to be key
\cite{TallonTallon-Bosc2008b,GillesWang2010a}. The poor performance of
the 3-step approach in these conditions is illustrated for instance in
Ramlau et al. \cite{RamlauEtAl2014a} with simulations of a
multi-object AO system at the ELT scale. The 3-step reconstructors
cannot be used with LGSs on the GSMTs.

\section{Kalman filter-based reconstructors}
\label{sec:kalman-filter-based}

The MCAO reconstructors presented above are known as static
reconstructors since they use only the information of the last
available measurement vector $\V{d}$.  The AO loop delay between the
time of the measurement and the time at which the AO correction is
applied has actually not been modeled in the equations above. The MCAO
reconstructors aims at compensating the atmospheric disturbances as
they were at the moment they were sensed. The AO correction is in
practice applied with some delay due to exposure time of the
sensor, reading or computation times. The static reconstructors do not
include any prediction to compensate for the fact that the
disturbances are no longer the ones that were measured. This is only
taken into account by the integrator control applying a gain on the
incremental command vector. This integrator gain has several purposes,
among which the aim to improve and stabilize the correction due to
this delay.

In order to make a prediction of what the phase distortions to correct
are, it is necessary to have a model of the temporal evolution of the
turbulent wavefronts. An example of such model is a first-order
auto-regressive model
\begin{equation}
  \label{eq:TemporalEq}
  \V{w}(k) = \M{T}_w \V{w}(k-1) + \V{v}(k)\,,
\end{equation}
where $k$ stands for the discrete time step, $\M{T}_w$ is the main
matrix of the model and $\V{v}$ is a zero-mean turbulence boiling
white noise term. The matrix $\M{T}_w$ is usually built based on the
spatio-temporal covariance of the wavefronts or on a spatio-temporal
linear interpolation of the wavefront modeling the frozen-flow
translations of the layers. The Kalman filter-based reconstructors
presented in this section differ from the reconstructors described
above by the fact that they rely on a temporal model equation similar
to Eq.~(\ref{eq:TemporalEq}). The system is modeled now by 2 equations
\begin{enumerate}
\item the measurement equation~(\ref{eq:MeasEq-CL}),
\item the temporal model equation~(\ref{eq:TemporalEq}).
\end{enumerate}

In the context of adaptive optics, the system is assumed linear and
the statistics of the turbulence and the noises are considered
Gaussian. The optimization of the MCAO criterion then fits into the
framework of Linear Quadratic Gaussian control (LQG). Such control
solutions for adaptive optics have been studied since the 1990's
\cite{PaschallAnderson1993a,Le-RouxRagazzoni2004a,KulcsarRaynaud2006a,PetitConan2009a,Looze2009a,MassioniKulcsar2011a}. It
is out of the scope of this paper to describe all possible approaches
to the implementation of an LQG control for MCAO (See
\cite{KulcsarEtAl2012a} and references therein). Their common
properties and characteristics are highlighted instead. In addition,
advantages and limitations of proposed LQGs solutions for AO are discussed.

A summary of the structure of Kalman filter based reconstructors for
MCAO is represented on the right part of
Fig.~\ref{fig:3stepR-and-KalmanR-schemes}. In this approach the
projection step is defined by the same equations as for the
minimum-variance reconstructors. The results of the projection is
directly applied to the DMs. There is no integrator-like control to be
inserted anymore before sending the commands to the actuators. This is
because the filtering and prediction are made in the Kalman estimator.

\subsection{Kalman estimator}
\label{sec:kalm-filt-equat}

The Kalman estimation includes the tomographic wavefront
reconstruction and the temporal prediction of the evolution of the
wavefront. This estimator uses the last available data, as well as
past data, and sometimes past control vectors applied to the DMs. The
computation of the Kalman estimate is made in 3 stages:
\begin{enumerate}
\item computation of the measurement error term (innovation term),
\item update of the state vector estimate (application of the Kalman gain),
\item prediction of the state vector to be compensated for (Kalman prediction).
\end{enumerate}
The first stage aims at comparing the newcoming measurements with what
the previous Kalman estimates predicts to be measured. If the
innovation term would be zero, it would mean that the last data do not
bring any new information that could not already be predicted by the
model. In such conditions, one would not need the measurement and
could base all the control of the AO correction on the predictive
model. Due to the random nature of the turbulence and its evolution,
as illustrated by the model in Eq.~(\ref{eq:TemporalEq}), the
innovation term is always informative. The second stage consists in
applying the Kalman filter gain on the innovation term, in order to
update the state vector estimate using the information brought by the
new measurements. The Kalman filter gain in AO is usually computed
from the asymptotic solution to a linear algebraic Riccati equation
\cite{KulcsarEtAl2012a}. The last stage consists in predicting the
state vector values, among which the tomographic wavefronts to be
corrected by the coming DM control vector.

\subsection{Benefits and limitations}
\label{sec:benefits-limitations}

The Kalman filter is by definition the optimal estimator and leads to
an optimal AO control. It can allow in principle an accurate modeling
of spatial and temporal properties of the AO system and of its
disturbances. Its predictive feature is a valuable asset, since
efficient prediction is expected to relax the constraints on the AO
loop frequency, translating into fainter limiting magnitudes for the
guide stars and greater sky coverage for the AO instruments.

Unfortunately, the implementation of such MCAO Kalman filter on a
real-time computer is still an open issue. In the same way the
minimum-variance reconstructors introduce some approximation of the
turbulence covariance $\Cprior$ to enable their fast implementation
(see Sect.~\ref{sec:impl-meth}), studied Kalman filters in the MCAO
literature also rely on some approximations in their models
\cite{KulcsarEtAl2012a,GillesMassioni2013a}. For instance, the
distributed Kalman filter (DKF)
\cite{MassioniKulcsar2011a,GillesMassioni2013a} uses a spatially
invariant approximation of the system in order to efficiently use fast
fourier transforms (FFTs) and fourier space operators of low
computational requirements. This prevents the filter to accurately
take into account the non-uniformity and the correlations of the noise
in the measurements from the lasers in case of elongated spots. Such
limitation has been shown to be prohibitive in
Sect.~\ref{sec:limitations-3-step} for use of the 3-step
reconstructors at the scale of the GSMTs. End-to-end simulations of
NFIRAOS system for the TMT \cite{GillesMassioni2013a} show in fact that
this crude approximation of the noise could only be compensated by the
predictive feature of the DKF reconstructor assuming a very good
knowledge of the wind speed atmospheric profile. It does not
outperform the minimum-variance FD-PCG reconstructor when uncertainties
on the atmospheric profile wind speeds orientation and amplitude are
larger than 20 degrees and 10\% respectively.

\section{Conclusions}
\label{sec:conclusions}

There is a large variety of MCAO reconstructors in the literature and
on the systems already working on-sky. Existing MCAO systems use
reconstructors based on interaction matrix inversion, using ad-hoc and
suboptimal regularizations. This comes from the fact that the MCAO
systems so far are handled in practice as if they would only be
extended classical AO systems.

MCAO correction can be optimized using minimum-variance
reconstructors. A large variety of fast algorithms have been studied
in the literature to afford the computation of minimum-variance
tomography in real-time at the scales of the GSMTs. The most
remarkable advantages of these reconstructors are
\begin{itemize}
\item adequate prior statistics using pseudo open-loop control
\item accurate noise modeling (in case of elongated spots)
\item ability to do global tomography or optimal split tomography between LGS and NGS data
\item optimization of the AO correction among the FOV using the projection step.
\end{itemize}

The importance of the accurate noise modeling for the GSMTs has been
highlighted comparing the minimum-variance reconstructors to the
3-step reconstructors and a Kalman filter based reconstructor, the
DKF, in which the noise model is approximated and considered
uncorrelated and uniform. For the ELT, the noise correlations prevent
the use of the 3-step reconstructors. For the TMT NFIRAOS system, the
DKF requires perfect wind speeds knowledge to compensate this crude
approximation by its predictive capabilities. Today, minimum-variance
reconstructors are the best solutions that can be implemented on sky
for large MCAO systems.

Kalman filter-based
reconstructors have a great potential for MCAO
\begin{itemize}
\item if the computational requirements can be reduced keeping at the same time reasonable approximations of the model
\item if atmopsheric turbulent layers wind speed can be accurately known.
\end{itemize}

\bibliography{ClemeBiblio} 
\bibliographystyle{spiebib} 

\end{document}